\documentclass[aps,pra, twocolumn]{revtex4-1}
\usepackage{amsthm, comment}
\usepackage[utf8]{inputenc}
\usepackage[sc,osf]{mathpazo}
\usepackage{graphicx,subfigure}
\usepackage{bm}
\usepackage{makeidx} 
\usepackage{amsmath}
\usepackage{amssymb}
\usepackage{color, xcolor}
\usepackage{amsthm}
\usepackage{hyperref}
\usepackage{cleveref}
\usepackage{diagbox}
\hypersetup{colorlinks, citecolor=blue, filecolor=blue, linkcolor=blue, urlcolor=blue}
\usepackage[normalem]{ulem}
\usepackage{float}
\usepackage{multirow}

\newcolumntype{P}[1]{>{\centering\arraybackslash}p{#1}}
\graphicspath{{Figures/}{Dynamics/}}
\begin{document}

\title{Predicting Critical Phases from Entanglement Dynamics in   $XXZ$ Alternating   Chain } 


\author{Keshav Das Agarwal, Leela Ganesh Chandra Lakkaraju, Aditi Sen(De)}

\affiliation{Harish-Chandra Research Institute, HBNI, Chhatnag Road, Jhunsi, Allahabad 211 019, India}

\begin{abstract}
 The quantum $XXZ$ spin model with alternating bond strengths under magnetic field has a rich equilibrium phase diagram which includes Haldane, Luttinger liquid, singlet, and paramagnetic phases. We show that the nearest neighbor bipartite and multipartite entanglement can detect quantum critical lines and phases in this model.    We determine a region in parameter space in which the dynamical states, starting from the  ground state of the Haldane (dimer) phase can create  highly multipartite entangled states for any time period, thereby establishing it as a potential candidate for the implementation of quantum information tasks.  We also exhibit that if the initial  and evolved states  are in two different phases, the nonanalytic behavior of multipartite entanglement and the rate function based on Loschmidt echo can signal quantum phase transition happened at zero temperature. In a similar spirit, we report that  from the product state, the patterns of  block entanglement entropy of the evolved state with time can also infer the phase transition at equilibrium. 
 

\end{abstract}

\maketitle

\section{Introduction}
\label{sec:intro}

\begin{figure}[h]
    \centering
    \includegraphics[width=\linewidth]{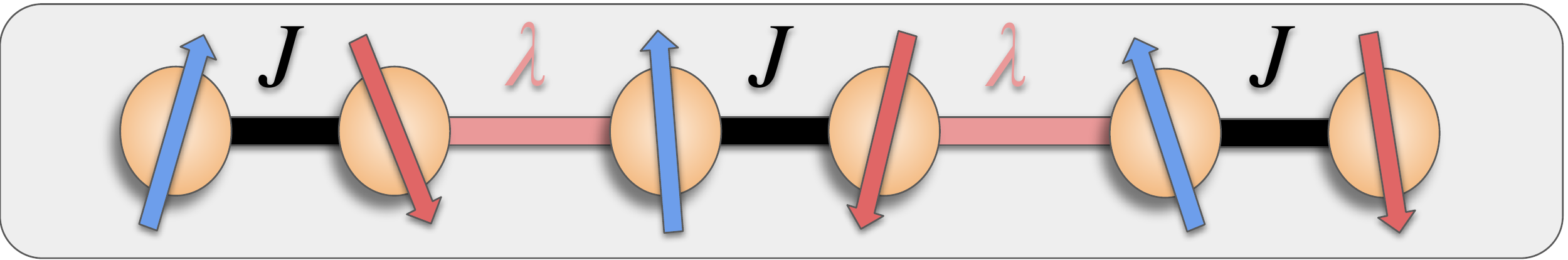}
    \caption{A schematic  diagram of an one-dimensional spin model consisting of the spin-$1/2$  particles governed by the Hamiltonian $\mathcal{H}^{alt}$, emphasizing on different bond strengths, \(J\) and \(\lambda\) between alternating pairs of particles.}
    \label{fig:schematic}
\end{figure}

At zero temperature,  quantum phase transition  occurs by tuning the magnetic field or the interaction strength in quantum spin models solely due to quantum fluctuations  \cite{qptbook2}. Hence, with the help of analytical, numerical and approximate methods, identifying these phases and critical lines via suitable physical quantities  is  important to increase the understanding of the system. In this respect, it was argued that some phases like Haldane phase \cite{haldane_original} can only be observed in spin models with integer spins  and not with half integer spins. However, it was  discovered that spin-$1/2$ Heisenberg antiferromagnetic-ferromagnetic (AF-F) alternating spin model in presence of magnetic field (see Fig. \ref{fig:schematic}) can show gapless Haldane phase  \cite{hida1,affleck_prl,hida_reference1,hida_reference2,phasediagram_3, sakai_reference1, sakai_reference2,sakai_reference3,sakai_reference4,sakai_reference5,sakai_reference6,sakai_reference7}. After the initial result,  numerical searches and variational methods reveal that this model possess a richer phase diagram which includes Luttinger liquid, Haldane, paramagnetic phases \cite{giamarchi2003quantum} compared to the corresponding antiferromagnetic or ferromagnetic Heisenberg spin chains \cite{original,phasediagram_1,phasediagram_2,new2,new3,new4}.  Although the static properties of this model has been  studied extensively, investigations on the  dynamical states are still missing which  may   reveal some counter-intuitive phenomena due to the competition between antiferromagnetic and ferromagnetic bonds. In this paper, we will show that it is indeed the case.

On the other hand, quantum spin models which can currently be simulated using cold atoms \cite{ultracold-review},  trapped ions \cite{ionrev} and superconducting qubits \cite{supcond,superc1,superc2}, turn out to be important test beds \cite{qpt,qpt1,qpt2,fazio-rev} for several quantum information processing tasks ranging from one-way quantum computation \cite{oneway1,oneway2,oneway3,oneway4,oneway5}, and quantum simulator \cite{simulator_lloyd} to quantum state transfer \cite{qst1}. In many of these processes, one requires  highly entangled states \cite{ent1} which are either the ground or  the thermal equilibrium  or the dynamical states of the spin models. 
It is interesting to identify  parameter regimes in the spin Hamiltonian which possess highly entangled states as the ground or  the thermal state below some critical temperature so that the model can be used as a resource of entanglement.  

Over the years, several mechanisms which include identifying physical observables have been developed to study the equilibrium physics of the many-body systems while such advancement is lacking  for the dynamical states \cite{Heylrev}.  From the perspective of information theory,  the investigation of evolution has two-fold motivations --  in one hand, determining initial states, quenching parameters and optimal time which can produce  highly entangled states  are useful for designing quantum protocols; on the other hand,  finding  quantum information theoretic quantities whose dynamical behavior  can faithfully reveal the criticalities at zero temperature, thereby mimicking the equilibrium physics can be important, both from the perspective of quantum information theory and condensed matter physics.


In this work, we are mainly interested to analyze the dynamical behavior of the $XXZ$ model with alternating bonds and magnetic fields, both from the perspective of answering fundamental questions as well as from assessing its capability as quantum devices. Before doing so, we first concentrate on the static properties  of this model. First, the critical lines between most of the phases, paramagnetic (PM), Luttinger liquid (LL), the combination of Haldane and singlet-dimer phases are drawn by studying the symmetry properties of the model \cite{original,phasediagram_1,phasediagram_2,new2,new3,new4}. 
We then employ bipartite and multipartite entanglement measures, Logarithmic negativity (LN) \cite{neg4, neg5} and generalized geometric measure (GGM) \cite{ggm_aditi,ggm1,ggm2,ggm3}, respectively, to distinguish between Haldane and singlet-dimer phase.   Furthermore, LN and GGM vanish in the PM phase while bipartite entanglement shows a decreasing nature in the LL  phase with the increase of the magnetic field, thereby  truly characterizing the phase diagram of this model.  We also report that bipartite entanglement at finite temperature can go beyond the entanglement content found in the zero-temperature state which can be important to probe the system in laboratories.  

 Our investigations on dynamics in the $XXZ$ model with alternating bonds are divided into three parts -- (1) creation of entanglement by appropriately driving the system parameters, (2) proposing multipartite entanglement as an indicator of dynamical quantum phase transition, (3) exhibiting the dynamics of block entanglement entropy which can signal quantum phase transition at zero temperature. 
In the $XXZ$ model with alternating bonds and magnetic fields, sudden quenching is performed either by tuning the anisotropy in the \(z\) direction or changing the interaction strength in the alternating bonds or both. We first identify a parameter space, both for the initial state and the evolution operator, which leads to a creation of high multipartite entanglement. Specifically, we observe that taking  ground states of Haldane or singlet phase as initials and  performing quench in these two phases, the evolved states created contain high multipartite entanglement. 

On the other hand, towards making connection between equilibrium and nonequilibrium physics,  it was realized that physical quantities like rate function based on Loschmidt echo (LE)  can show  nonanalyticity with time when the phase of the initial and the driving Hamiltonian are chosen from two different phases, thereby detecting a phenomenon known as dynamical quantum phase transition \cite{heyl_original,Heylrev}. 
We show here that both the rate function as well as GGM of the evolved state show nonanalyticities with time  when the system is initially in the ground state of the Haldane phase and finally quenched in the singlet-dimer phase and vice-versa.  
Moreover, if the initial state is prepared in a suitable product state, the patterns of block entropy in the evolved state can determine whether parameters in the Hamiltonian responsible for the evolution is close to the critical line or far from it. Specifically, we observe that in the transient regime, the block entanglement entropy  remains almost constant with time for different system parameters which are chosen from the Haldane or singlet phase and far from the Haldane-singlet critical line while the entropy  behave differently with the variation of time when the parameters in the Hamiltonian are chosen close to the Haldane-singlet phase boundary. 

The paper is organized in the following way. The $XXZ$ model with alternating bonds under investigation  and its symmetry properties are described in Sec. \ref{sec:Ham_Sym} while the detection of critical lines via bipartite as well as multipartite entanglement is reported in Sec. \ref{sec:staticent}.   Sec. \ref{sec:entdyn}  identifies a parameter regime and time period in which maximal genuine multipartite entanglement can be generated.  In Sec. \ref{sec:dqpt}, multipartite entanglement as well as entanglement entropy obtained in the  dynamical states are used to characterize quantum phase transitions at zero temperature. We present concluding remarks in Sec. \ref{sec:conclu}.

\section{$XXZ$ Alternating Chain and its symmetries}
\label{sec:Ham_Sym}

\begin{figure}[H]
		\centering
		\includegraphics[width=7.5cm]{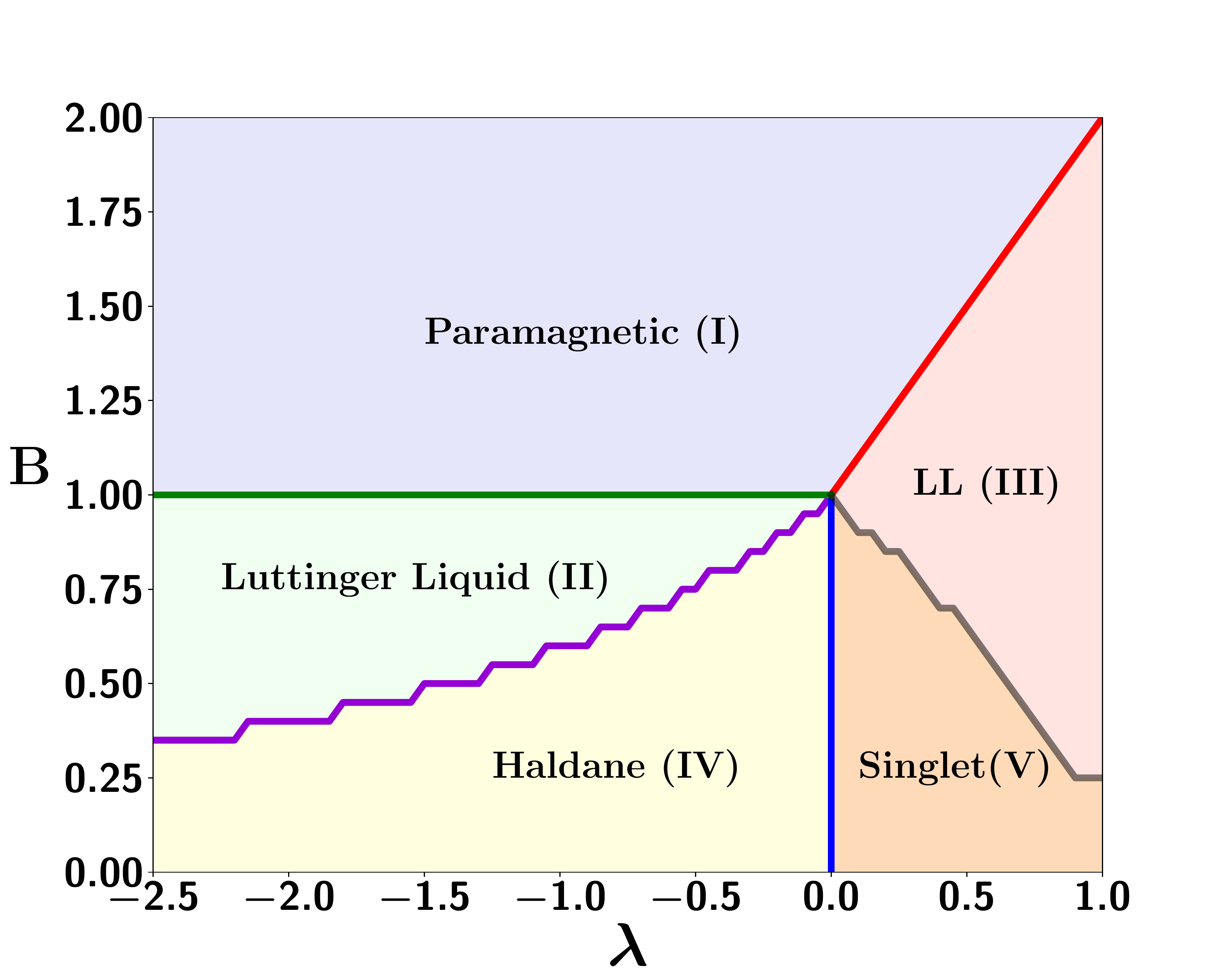}
		\caption{ \textbf{Phase diagram of the $XXZ$ model with alternating bonds \cite{original}.} The critical lines are plotted in the \(\lambda, B\)-plane by considering the energy gap between the two lowest eigenstates (see text for details). 
		Here $\Delta~=~1$ and \(N=16\). The phases, paramagnetic (PM) (I), Luttinger liquid (LL) ((II) and (III)), Haldane (IV) and singlet-dimer (V) are mentioned according to the literature (see Refs. \cite{original,phasediagram_1,phasediagram_2,new2,new3,new4} and reference therein).  }
		\label{fig:HacMf_phase_diagram}
	\end{figure}

Let us describe the properties of the
$XXZ$ model with the alternating bond strengths and magnetic field which admits a richer phase diagram than the Heisenberg chain \cite{original,phasediagram_3}. The Hamiltonian of the model reads as
\begin{eqnarray}
\label{eq:Hacmf}
	\mathcal{H}^{alt} & = & J \sum_{i=1}^{N/2} (S_{2i-1}^x S_{2i}^x + S_{2i-1}^y S_{2i}^y + \Delta S_{2i-1}^z S_{2i}^z)  \nonumber \\
	& + & \lambda' \sum_{i=1}^{N/2} (S_{2i}^x S_{2i+1}^x + S_{2i}^y S_{2i+1}^y + \Delta S_{2i}^z S_{2i+1}^z) \nonumber \\
	&- & B' \sum_{i=1}^{N}S_i^z, 
	\end{eqnarray}
where $S^k_i$ (\(k=x, y, z\)) represents the  spin operators (which is half of the Pauli spin operator) at site $i$, in the chain of length \(N\), the coupling parameters of odd and even bonds are respectively $J>0$, indicating  antiferromagnetic interaction strength and $\lambda$ which can be both positive and negative, \(\Delta >0\) denotes the anisotropy parameter in the \(z\) direction and \(B'\) is the strength of the magnetic field. We set \(\lambda = \lambda'/J\) and \(B = B'/J\) throughout the paper. We also assume the periodic boundary condition, i.e., \(S_{N+1} = S_1\) with \(N\) being the even.  Notice that it permits a regime where even bonds are antiferromagnetic while odd bonds are ferromagnetic, as well as both the bonds are antiferromagnetic although the strength can be different in alternating bonds, thereby making both the systems nondiagonalizable analytically. We refer to the former one as antiferromagnetic-ferromagnetic $XXZ$  alternating chain and the latter one as antiferromagnetic $XXZ$ chain with alternating bond strengths.  The numerical techniques like Lanczos algorithm \cite{Lanczos, dmrg_white}, and approximate procedures are employed to investigate  the phase diagram via the energy gap, spin correlation functions of this model.     

It was found that the competition between antiferromagnetic and ferromagnetic bonds lead to  a rich quantum phase diagram  in the $(B,\lambda)$-plane \cite{original,phasediagram_3} which include Luttinger liquid  (LL), Haldane, paramagnetic (PM) phases.  When both the bonds are AF although the strengths in alternating bonds are different,  it also possess singlet-dimer, LL and PM phases with the changes of the magnetic fields and \(\lambda\) (as shown in Fig. \ref{fig:HacMf_phase_diagram}). 
In this paper, one of the primary goals is to  investigate the dynamical state, and so we use Lanczos algorithm for diagonalizing the Hamiltonian \cite{Lanczos}. This method first converts the sparse-matrix Hamiltonian into a tridiagonal matrix in the Krylov subspace basis which makes the diagonalization problem simpler.  As mentioned next, we also exploit some symmetry properties of the system  so that the higher number of total spins  especially during dynamics can be addressed.  

As depicted in Fig \ref{fig:HacMf_phase_diagram},  we reproduce the phase diagram  of the Hamiltonian \(\mathcal{H}^{alt}\) with \(\Delta =1.0\) \cite{phasediagram, phasediagram_1, phasediagram_2} by computing the energy gap, denoted as \(\Delta E\),  which is the difference between the two lowest eigenenergies. Except the transition from the Haldane (denoted as (IV)) to the singlet-dimer phase  (mentioned as (V)), we draw the critical lines  when the energy gap is less than \(0.05\) for \(N=16\) which are in a good agreement with previously known results.  However, the corresponding phases are mentioned according to the literature \cite{phasediagram, phasediagram_1, phasediagram_2}.  We will also show in succeeding section that entanglement can detect all the critical lines in this model.


\subsection{Ground state symmetry properties}
\begin{figure*}[ht]
		\centering
		\includegraphics[width=\linewidth]{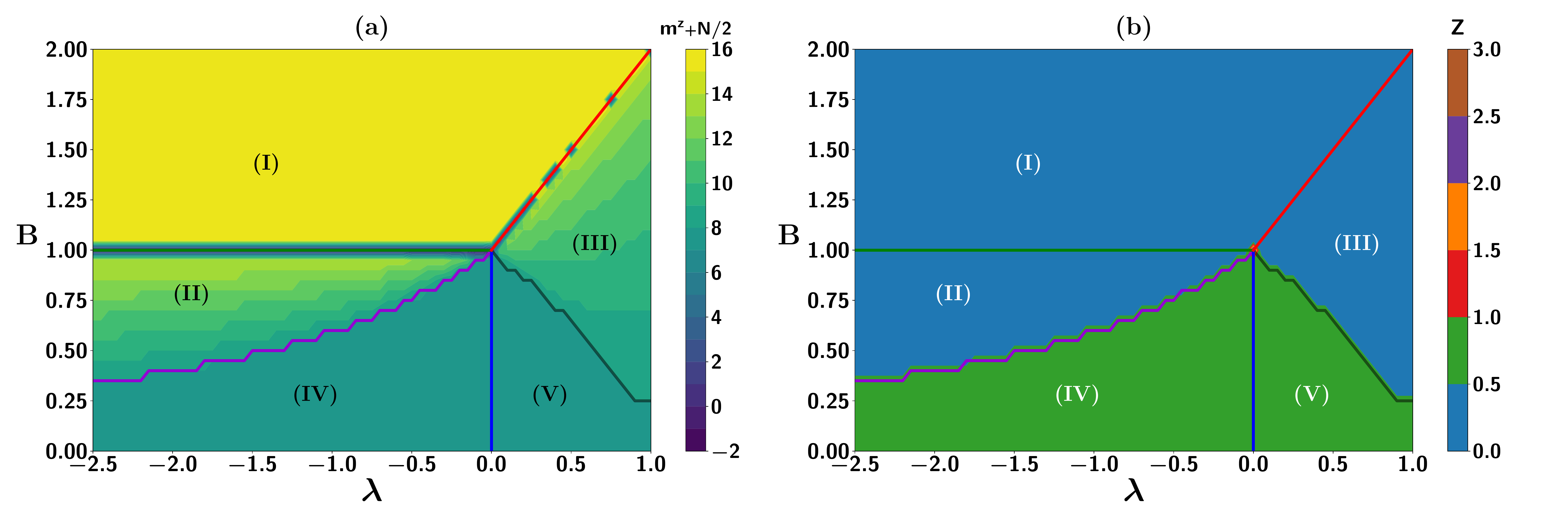}
		\caption{\textbf{Symmetries of the ground state.}  (a) Map plot of the total magnetization in the \(z\) direction, $m^z + N/2$, of the ground state with respect to \(\lambda\) (horizontal axis)  and \(B\) (vertical axis).  (b) Map plot of spin inversion  operator ($Z$) defined in Eq. (\ref{symmetries}) in the \(\lambda,B\)-plane. Here the system-size is taken as \(N=16\). Magnetization can distinguish all the phases except (IV) and (V) while the spin inversion operator can distinguish  (I)-(III) together with (IV)-(V). For degenerate eigenstates, we use a convention  which is mentioned in the text.  Both the axes are dimensionless.  }
		\label{fig:gs_sym}
	\end{figure*}

The Hamiltonian in Eq. (\ref{eq:Hacmf}) possess various symmetries -- translation symmetry (\(T\)), parity (\(P\)) and spin inversion (\(Z\)) for $\sum_{i=1}^{N}S_i^z=0$ \cite{Sanvik_symmtery}. In the computational ($S^z$) basis, they are represented by
\begin{equation}
\begin{split}
	T|S^z_0, S^z_1,S^z_2....S^z_N\rangle &= |S^z_2....S^z_N,S^z_0, S^z_1\rangle,\\
	P|S^z_0, S^z_1,S^z_2....S^z_N\rangle &= |S^z_N,...... S^z_2,S^z_2,S^z_0\rangle,\\
	Z|S^z_0, S^z_1,S^z_2....S^z_N\rangle &= |-S^z_0, -S^z_1,-S^z_2....-S^z_N\rangle.
	\end{split}\label{symmetries}
	\end{equation}
These symmetries admit certain good quantum numbers which can be used to label the basis states. Due to the translational symmetry, the states with momentum can be represented as $k\in\{0,1,....N/4\}$, the parity leads to $p\in\{1,-1\}$ and the spin inversion gives  $z\in\{1,-1\}$. Interestingly, $k$ becomes semi-momentum when the parity symmetry is applied, which is $\frac{2\pi k}{N/2}$. The Hamiltonian in Eq. (\ref{eq:Hacmf}) also preserves the total spin in the \(z\)-direction, and  its quantum number is represented by $m^z = \sum_{i=1}^{N}S_i^z$. Therefore, $m^z\in\{-N/2,-N/2+1, \cdots, N/2-1,N/2\}$ for the ground state.  The behaviors of \(m^z\) and spin inversion numbers are shown in Fig. \ref{fig:gs_sym}, from which one can also infer the phases  or critical lines of the model. 

	
\begin{enumerate}
    \item \emph{Total spin magnetization in the \(z\) direction.} We are interested in the trends of  $m^z+N/2 \in \{0,1,...N\}$ considering the ground state.  If there is two-fold degeneracy and $m^z$ is different for the two lowest energy states,  $m^z+N/2$ is set to $-1$ while for higher degeneracy, it is taken to be $-2$. Notice that the blue line differentiating LL with the PM phases has two-fold degeneracy and the two eigenstates correspond to two different \(m^z\) values. 
    
    In the Haldane as well as the singlet-dimer phases, $m^z+N/2$ is $N/2$, thereby unable to distinguish them while it is $N$ in the paramagnetic phase, suggesting that it contains the spins, polarized in the up-direction (see Fig. \ref{fig:gs_sym}(a)), which can also be confirmed by nearest-neighbor correlations in the next section. 
	Since the Lutinger liquid phase has degenerate eigenstates, it has no definite value and it increases with the increase of the magnetic field, \(B\). 
	Therefore, the behavior of \(m^z\) can differentiate the regions between paramagnetic, Luttinger liquid and the Haldane-singlet together. 

	\item \emph{Semi-momentum.} The states with semi-momentum number, \(k\), can differentiate only between LL in the AF-F alternating chain and the rest since in all other phases, it vanishes for the ground state. Like \(m^z\), again the critical line between LL and PM phases (between (I) and (II)), there is two-fold degeneracy with different \(k\) number and hence we use same convension as \(m^z\).   
   
	
	\item \emph{Parity.} Unlike semi-momentum, the parity operator, \(p\) is nonvanishing only when the ground state is in the Luttinger liquid phase of the AF Heisenberg alternating chain. Again, in the degenerate ground state space (between (I) and (III)), we use the similar fixed value as in \(m^z\). 
	
%
	\item \emph{Spin inversion.} The valid $z$ values are  $0, 1$ and $-1$. It can be seen that $z$ can be used to differentiate between every other phases and the Haldane-singlet combination as shown in Fig. \ref{fig:gs_sym}(d). Since there is no spin inversion symmetry in phases (I), (II) and (III), we put zero in the computation. Note that the critical point where all the phases meet has higher degeneracy. 
	
\end{enumerate}

From the collective analysis of all the symmetries, we can infer the paramagnetic (I),  Luttinger liquid in the AF-F chain (II), and LL (III) phase in the AF chain although the Haldane and singlet phases cannot be differentiated. We will show in the succeeding section that patterns of entanglement is capable to distinguish the Haldane phase from the singlet one. 




\section{Detecting criticalities via entanglement}
\label{sec:staticent}

In this section, we will argue that bipartite as well as multipartite entanglement measures of the zero-temperature (ground) states are capable to identify quantum critical lines and they  behave in a distinct manner in each phases, thereby showing their recognizing power in this model. This study also plays an important role to establish that the trends of entanglement measures of the evolved state can also mimic the phase diagram observed in the ground state.  

Before  investigation of entanglement, let us first compute the energy gap, \(\Delta E\), important for studying entanglement and nearest neighbor classical correlator in the \(z\) direction. We also report the nonmonotonic behavior of bipartite entanglement for the canonical equilibrium state of this model. The entire study has been carried out by exactly diagonalizing the Hamiltonian using Lanczos method and the symmetry properties mentioned in the previous section. All the analysis at the zero temperature is performed for \(N=16\) while in case of the thermal and dynamical states, we consider the system size to be \(12\). 



\begin{figure}[h]
		\centering
		\includegraphics[width=9.5cm]{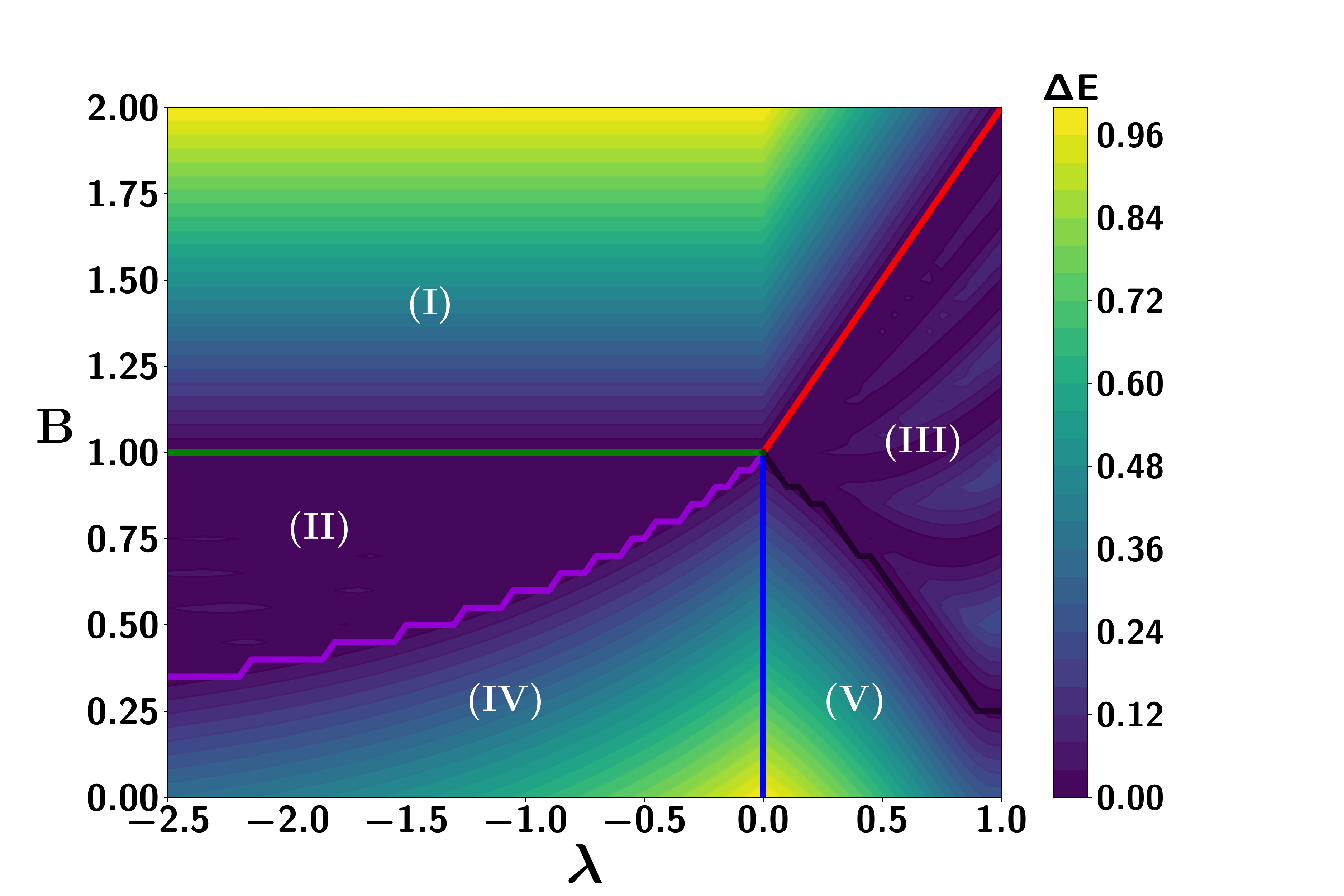}
		\caption{\textbf{Energy gap in different phases.} Map plot of energy gap between the ground and the first excited state against \(\lambda\) (abscissa) and \(B\) (ordinate). System size is same as in Fig. \ref{fig:HacMf_phase_diagram}. Both the axes are dimensionless. }
		\label{fig:Ediff}
	\end{figure}

\textbf{Energy gap.} Let us first study its behavior in the AF-F alternating spin chain. In the Haldane phase, \(\Delta E\)  increases with the increase of \(\lambda\) for a fixed value of \(B\) while in the PM phase, it remains constant with \(\lambda\) although it starts increasing with the increase of \(B\) as depicted in Fig. \ref{fig:Ediff}.  In the LL phase, it vanishes, thereby suggesting the gapless phase \cite{luttinger}. We will show that the information about gapless-gapped phases can be important for computing bipartite as well as multipartite entanglement.

On the other hand, in the AF $XXZ$ model with alternating bond strength, the energy gap decreases with the increase of \(\lambda\) for a given magnatic field, \(B\) in the singlet-dimer phase while in the PM phase, it increases with the variation of \(B\) although it decreases with the increasing \(\lambda\) values as shown in Fig. \ref{fig:Ediff}.  
Moreover, some regions of LL phase and the phase boundary between the LL and the PM phase show vanishing energy gap.


\textbf{Nearest neighbor spin correlation function.} Since we find that the ground state is degenerate in the LL phase, to compute any nearest neighbor physical quantities, we consider the canonical equilibrium state of the Hamiltonian with a very low temperature, which reads  at temperature \(T\) as  \(\rho(\beta) = \frac{\exp(-\beta \mathcal{H}^{alt})}{Z}\) where \(Z = \text{Tr}(\exp(-\beta \mathcal{H}^{alt}))\) is the partition function with \(\beta = 1/K_B T\), \(K_B\) being the Boltzmann constant. To distinguish it from the ground states, we call it as the zero-temperature state. 

To compute any two-party observables, the two-party reduced density matrix, \(\rho_{ij}(\beta)\) can be obtained from \(\rho(\beta)\) by tracing out all the parties except sites \(i\) and \(j\). 
 Note that the Hamiltonian has a double-translational invariance due to the presence of the alternating bond strengths $J$ and $\lambda$, i.e., the Hamiltonian remains  invariant, when  spins are shifted twice. Therefore, two nontrivial nearest neighbor density matrices, namely $\rho_{i(i+1)}$, ${i=1,2}$ corresponding to $i \in \{\text{odd}, \text{even}\}$ have to be studied. 

\begin{figure}[ht]
		\centering
		\includegraphics[width=9.5cm]{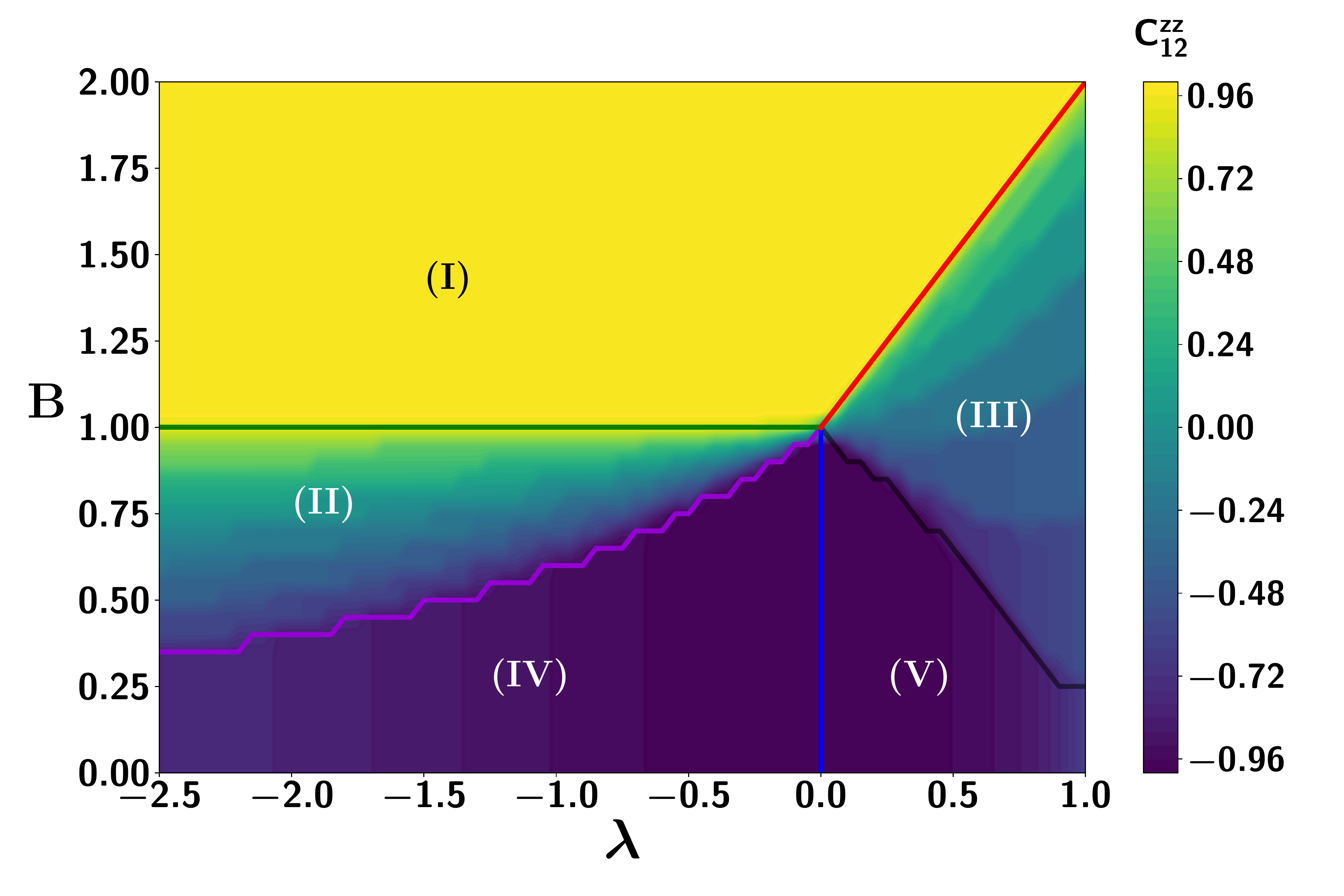}
		\caption{Map plot of the classical correlators, $C^{zz}_{12}$ for nearest neighbor sites of the ground state. All other specifications are same as in Fig. \ref{fig:Ediff}.}
		\label{fig:cz_gs}
	\end{figure}

We now compute two nearest neighbor classical correlators in the \(z\) direction, given by $C^{zz}_{ij} = Tr[\rho_{ij} \sigma^z_i\otimes\sigma^z_{j}]$ for the zero-temperature state where \(\rho_{ij}\) is obtained from the thermal state of the Hamiltonian with \(\beta =10^4\). Firstly, 
in the PM phase, the values of the nearest neighbor classical correlator in the \(z\) direction is maximum for both $i=\{1,2\}$ and remains constant with the variation of \(\lambda\) as well as \(B\) while there is a variation from negative to positive in the LL phase (see Fig. \ref{fig:cz_gs}).   On the other hand, when \(\lambda\) increases,  $C^{zz}_{12}$  decreases  in the Haldane phase while it  increases in the singlet-dimer phase with \(\lambda\), thereby distinguishing these two phases which is not possible by using symmetries of this model. Note, however, that the change with \(\lambda\) (increase or decrease) is very small, which is almost unnoticeable.  We will show that  such features become  more pronounced when bipartite entanglement is calculated in the next subsection. Furthermore, \(C_{23}^{zz}\) does not show any additional  feature which cannot be  seen in  \(C_{12}^{zz}\). 


\subsection{An identifier of criticalities: Multipartite vs. bipartite entanglement}


\begin{figure*}[ht]
		\centering
		\includegraphics[width = \linewidth]{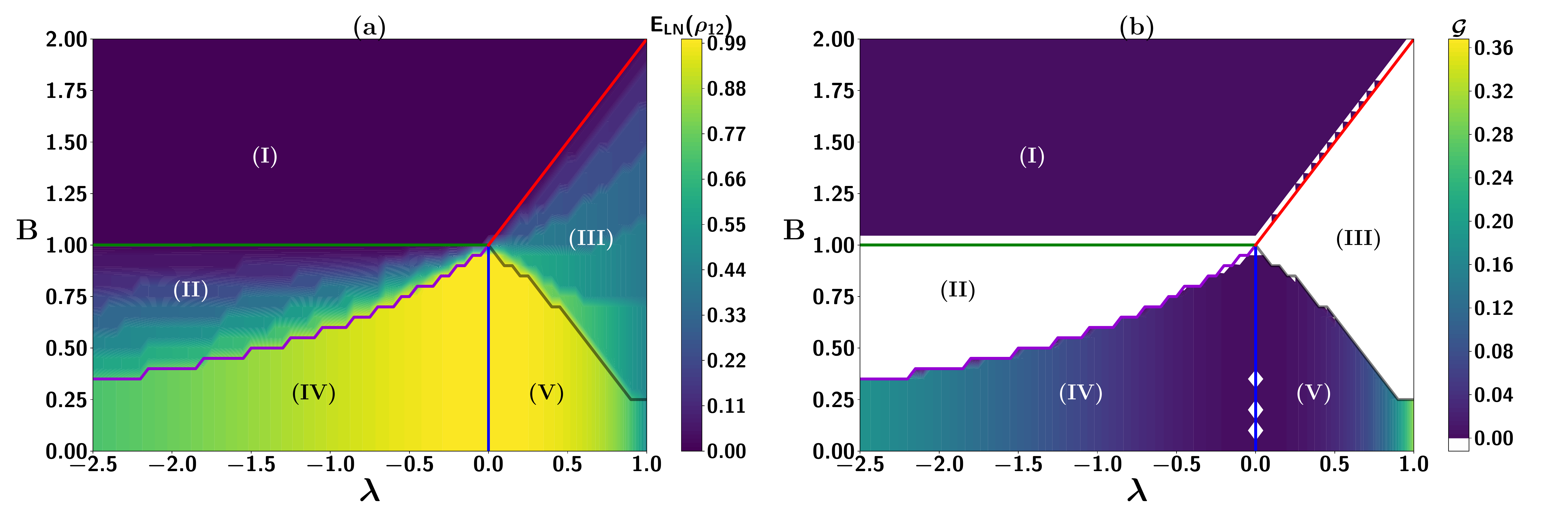}
		\caption{\textbf{Entanglement in the ground state of the $XXZ$ model with alternating bonds.} (a) Map plot of nearest-neighbor  entanglement, $E_{LN}(\rho_{12})$ for the zero-temperature state in the \((\lambda,B)\)-plane. (b) GGM, $\mathcal{G}$ for the ground state in the same parameter space. As the Luttinger liquid phase corresponds to degenerate ground states, GGM is not computed. The corresponding color palettes display the value of the $E_{LN}(\rho_{12})$ and $\mathcal{G}$ respectively. Interestingly, $E_{LN}(\rho_{12})$ decreases with the increase of \(|\lambda|\) in the Haldane as well as singlet-dimer phase while the opposite picture emerges for GGM due to monogamy of entanglement. Here \(N=16\). Both the axes are dimensionless.  }
		\label{fig:lneg_gs}
	\end{figure*}

Let us first present the bipartite and multipartite entanglement measures used to investigate the behavior of the static and dynamical properties of the system. Bipartite entanglement is quantified by logarithmic negativity (LN) \cite{neg4,  neg5}, given by
\begin{equation}
\label{lneg_eq}
	E_{LN}(\rho_{ij}) = \log_2||\rho_{ij}^{T_i}||
	\end{equation}
where $\rho_{ij}^{T_i}$ denotes the partial transposition over the site $i$ \cite{peres}. Since we are dealing with two spin-1/2 particles, \(E_{LN} (\rho_{ij})\) reduces to the modulus of the negative eigenvalue of the partial transposed state. In this scenario, \(\rho_{ij}\) is either \(\rho_{12}\) or \(\rho_{23}\) obtained from the thermal state with high \(\beta\).  

We are also interested to find out the behavior of genuine multipartite   entanglement in the ground state. A pure state, \(|\psi_N\rangle\) is said to be genuine multipartite entangled if it is not product in any bipartition. A distance-based measure, called generalized geometric measure (GGM), can quantify the genuine multipartite entanglement content present in the \(N\)-party state \cite{ggm_aditi,ggm1,ggm2,ggm3} which is  defined as 
\begin{eqnarray}
     \mathcal{G}(|\psi_N \rangle) & = & 1 - \max_{\{|\phi_{nong}\rangle\}} |\langle \phi_{nong} | \psi_N\rangle|^2 \nonumber\\
     & = & 1- \max_{i}\{(\lambda^m_{i:rest})^2\},
    \label{GGm_def}
\end{eqnarray}
where maximization in the first line is performed over the set of all nongenuinely multipartite entangled states, \(\{|\phi_{nong}\rangle\}\) while in the second line, the maximization is  over the set of all the maximal Schmidt coefficients obtained from all  possible bipartitions of \(|\psi_N\rangle\),  \(\{(\lambda^m_{i:rest})^2\}\). Since we will dealing with total number sites to be  sixteen or twelve, calculating \(\mathcal{G}\) after maximizing over all eigenvalues from all bipartitions is computationally costly and hence we restrict to  \(\mathcal{G}\) which considers all single and two-site reduced density matrices. Note that this restricted GGM is also a multipartite entanglement measure. Furthermore, numerical searches for \(N=8\) and \(N=10\) reveal that in this model, the actual GGM coincides  with the restricted one except at the boundary of Haldane-singlet phase. Hence, it is reasonable to assume that all the calculations presented in this paper is indeed quantifying genuine multipartite entanglement.  





\textbf{Trends of LN and GGM.} We start our investigation by noting that the paramagnetic state is  all up-spin, which is separable at every bipartition, and  therefore, both LN and GGM vanish in this phase. Since LL phase is degenerate, GGM cannot be calculated (since it is hard to calculate for mixed state (cf. \cite{aditi_ggm_mixed,ggm_mixed_otfried})) although $E_{LN}(\rho_{12})$ is significant  in  the LL phase and decreases with the increase of \(B\) as shown in Fig. \ref{fig:lneg_gs}(a). 

In a multipartite domain, monogamy of entanglement \cite{ckw, ckw2, neg-sq} puts restriction on the shareability of entanglement between pairs, i.e., it  says  in a \(N\)-party state, if two of them are highly entangled, the other pairs possess a  less amount of entanglement.   In this respect, the contrasting features emerge in the Haldane-dimer phase (comparing Figs. \ref{fig:lneg_gs}(a) and (b))  -- in the Haldane phase, $E_{LN}(\rho_{12})$  increases with the increase of \(\lambda\)  while the decreasing nature of GGM can be observed in this phase. The opposite picture is seen in the singlet-dimer phase, i.e., GGM increases and LN decreases with the increase of \(\lambda\). It can be argued that   the competition between the interaction strength of  even and odd bonds leads to a high multipartite entanglement in a system with high \(|\lambda|\).  
It also manifests that if we are interested to utilize bipartite entanglement of the zero-temperature state, the region close to the transition of Haldane and singlet-dimer phases is favorable while in case of multipartite entanglement, far from the transition region of Haldane-singlet criticalities possess  highly entangled states. 


\emph{Nonmonotonicity of entanglement with temperature.} It has been observed in almost all quantum spin models that nearest neighbor  entanglement of the thermal state with \(\beta >100\) mimic the entanglement of the ground state.  Similarly, entanglement vanishes when \(\beta\) is sufficiently low. The question is whether entanglement goes to zero monotonically with the variation of  \(\beta\) or not. It was  reported for several one-dimensional spin models that such monotonic behavior does not hold  \cite{nonmontonicity1,nonmontonicity2,nonmontonicity3,nonmontonicity3,nonmontonicity4,nonmontonicity4,nonmontonicity5,nonmontonicity6,nonmontonicity7,nonmontonicity8,nonmontonicity9}. The spin-1/2 $XXZ$ model with alternating bonds has a rich phase diagram and so it will be interesting to see whether any nonomonotonicty appears in this case also. To check that, we compute
$\beta^{max}_{ij} = \text{argmax}_{\beta}[E_{LN}(\rho_{ij}(\beta))]$ with $(i,j) = (1,2)$ and $(2,3)$,  i.e., we look for \(0< \beta \leq 100\)  which can possess maximum nearest neighbor entanglement.    On varying the inverse temperature, $\beta$,  $E_{LN}(\rho_{ij}(\beta))$ is maximised.  

\begin{figure}[ht]
		\centering
		\includegraphics[width=\linewidth]{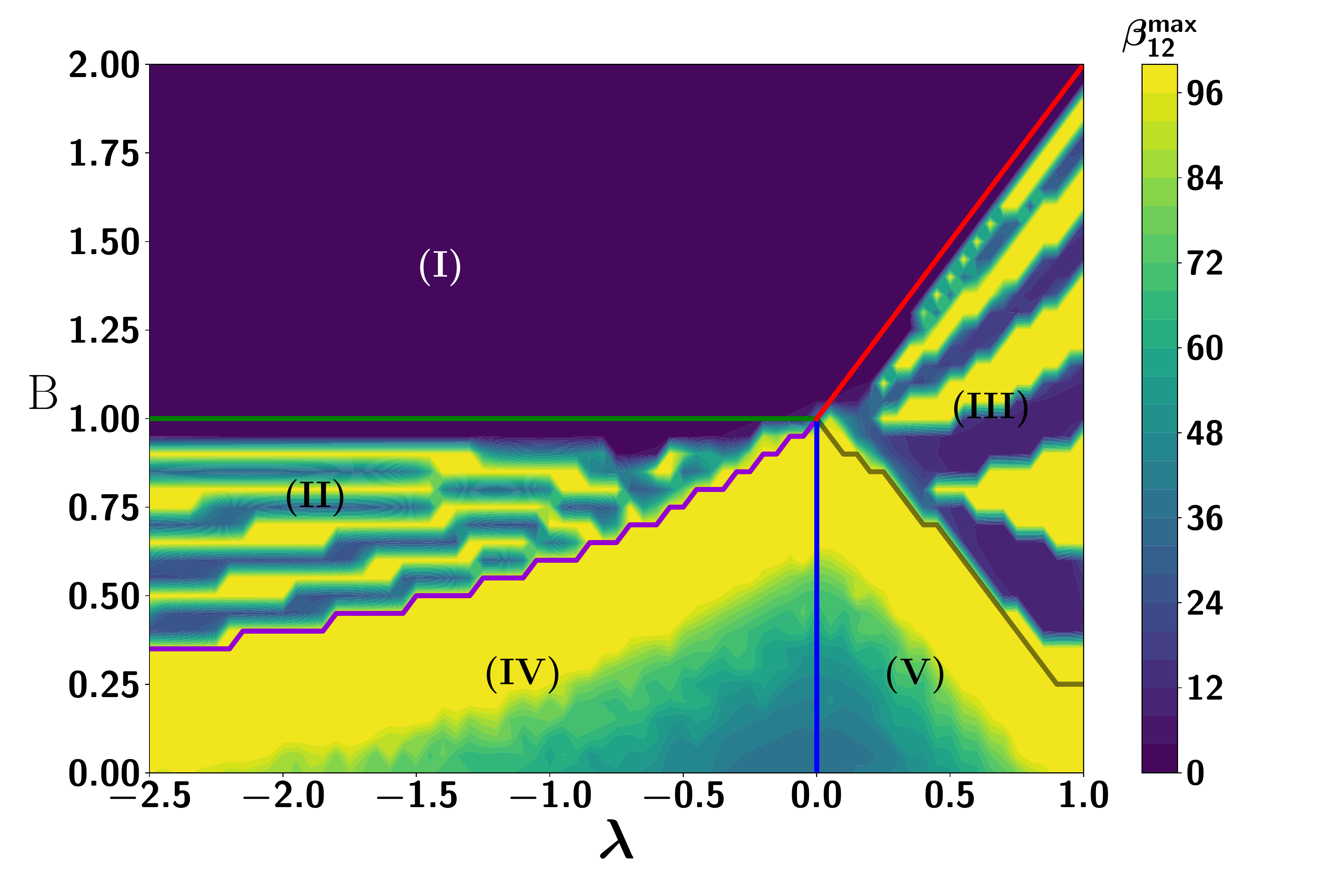}
		\caption{Contour plot of the inverse temperature ($\beta_{12}^{max}$) corresponding to the maximum entanglement of $\rho_{12}$ in the  $B$  (vertical axis) and $\lambda$ (horizontal axis) plane. The color palette displays the values of  $\beta_{12}^{max}$. It shows that there exists a finite temperature which possess a high bipartite entanglement compared to the state at zero-temperature. Here \(N=12\). Both the axes are dimensionless.  }
		\label{fig:bmax}
	\end{figure}

The observations emerge from  Fig. \ref{fig:bmax} are as follows: (a) we see that in the Haldane and singlet phase, $\beta_{12}^{max}$ shows non-monotonic behavior in the neighborhood of $\lambda \in \{-1.5,0.5\}$, where the corresponding maximum entanglement  is also high in the Haldane and the singlet-dimer phases. Similar nonmonotonic behavior is also observed in the LL phase. 

\section{Production of high Multipartite entanglement in dynamics}
\label{sec:entdyn}

In the literature, there exist quantum spin models like Ising model  which can produce maximal  multipartite entanglement for a certain time period with a suitable initial state \cite{oneway1,oneway2,oneway3,oneway4,oneway5}. As discussed in the preceding sections, since the considered $XXZ$ model with alternating bonds contain rich and exotic phases, we expect that its dynamical state can also have capability to generate  highly  entangled states. Indeed we find that it is true. 

\begin{figure*}[ht]
		\centering
		\includegraphics[width=\linewidth]{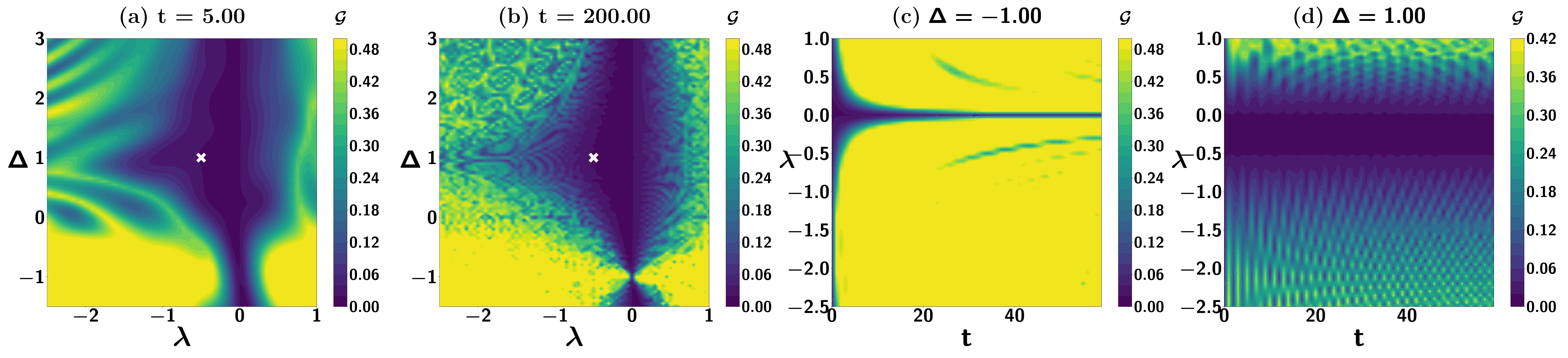}
		\caption{\textbf{Creation of entanglement in dynamics of the $XXZ$ with alternating bonds.}
		(a) Contour plot of GGM, $\mathcal{G}$ for the evolved  state at time $t = 5.00$. The system is initially  in the ground state of the  Haldane phase, i.e., the ground state of  $\mathcal{H}^{alt}(\lambda_0 = -0.5, B_0 =0.25, \Delta_0 = 1,)$ is the initial state marked by the cross.  The values of \(\lambda\) (abscissa) and \(\delta\) (ordinate) represent the  sudden quench  of the Hamiltonian at \(t>0\). (b) Same as in (a) with $t = 200$.  (c) Time evolution of $\mathcal{G}(t)$ against $\lambda$ (values after the sudden quench at \(t>0\)) (vertical axis) and \(t\) (horizontal axis) with a fixed quench \(\Delta =-1.0\). The initial state is same as in (a).  (d) Similar plot as (c) except no quenching in \(\Delta\). The color palette in all the four figures correspond to the values of $\mathcal{G}$. The entire analysis is performed with the system-size, \(N=12\).  All the axes are dimensionless. }
		\label{haldane_t_delta}
	\end{figure*}

Since we are interested in the computation of GGM, the initial state cannot be chosen from the ground state with PM and LL phases since the GGM of the former vanishes and in the latter case, the ground state space is degenerate. The entire analysis in dynamics is performed when the initial state is either chosen as the ground state in the Haldane or in the singlet-dimer phase. Let us denote the initial state  at \(t=0\) as \(|\psi_{0}\rangle\) which is prepared in the ground state  of \(\mathcal{H}^{alt}_0 \equiv \mathcal{H}^{alt}(\lambda_0, \Delta_0, B_0)\).  At \(t>0\), the sudden quench is performed by changing \(\lambda_0\) to \(\lambda\) or \(\Delta_0\) to \(\Delta\) or both. In this study,  sudden quench in the magnetic field is not executed since $B$ has no effect on the quenching. This can be easily explained by the  commutator of $\mathcal{H}^{alt}_0$ and $\mathcal{H}^{alt}_1$ at \(t=0\) and \(t>0\) respectively given by 
\begin{equation}
    [\mathcal{H}^{alt}_1,\mathcal{H}^{alt}_0]= i (\lambda-\lambda_0)\sum_{j=1}^N (-1)^j \epsilon_{pqr}(S_j^p S_{j+1}^q S_{j+2}^r)
\end{equation}
where \(\lambda_0\) and \(\lambda\) represent the initial and final interaction strengths,  $p,q,r \in {x,y,z}$, $\epsilon_{pqr}$ is the Levi-Civita symbol, \(S_{N+i} = S_i\), \(\forall i\). Since the commutator is independent of the initial and the  final magnetic fields,  quenching with same $\lambda$ and different $B$ gives the same eigenvector as initial and final, thereby leading to  a trivial evolution, 

Therefore, the evolved state at time (\(t>0\)) can be written as 
\begin{equation}
    |\psi_{t}\rangle = \exp(-i \mathcal{H}^{alt}(\lambda, \Delta, B) t) |\psi_{0} (\lambda_0, \Delta_0, B_0)\rangle. 
    \label{eq:quench}
\end{equation}
%
%
%
As shown in Fig. \ref{fig:lneg_gs}(b), GGM  of the ground state is very low in the neighborhood of the Haldane-singlet transition line. Hence, it is reasonable to determine whether entanglement generation is possible in this domain. Towards this aim, we prepare the initial state in the Haldane phase and for illustration, we choose the initial system parameters to be   $B_0=0.25$,  $\lambda_0 = -0.5$ and \(\Delta_0 =1.0\). At this point, $\mathcal{G}(|\psi_0\rangle) = 0.0189$.  Evolving the state with the unitary operator $U (\Delta, \lambda, B)=e^{-iH(\Delta, \lambda, B)t}$, and computing  \(\mathcal{G}(|\psi_t\rangle)\) by varying $t$, we observe the following characteristics:
\begin{enumerate}
    \item \emph{Creation of maximal GGM states.} We find that starting from the initial state in the Haldane phase,  at \(t>0\), any sudden quench of \(\lambda\) in the Haldane or in the singlet phase along with the sudden change of \(\Delta\) from \(1.0\) to \(-1.0\) leads to a high amount of genuine multipartite entanglement generation over the entire time period except the critical line between the Haldane and the singlet phase as depicted in Fig. \ref{haldane_t_delta} (c). Specifically, we observe that  in this situation, the evolution can create  almost maximally genuine multipartite entangled state, in the entire time duration, (see Fig. \ref{haldane_t_delta}(c)) i.e., 
    \begin{eqnarray}
         &&\max \mathcal{G}(|\psi_t\rangle ) = \max \mathcal{G}(e^{-i\mathcal{H}^{alt}t}|\psi_0\rangle) \approx 0.5
         \, \, \text{with} \nonumber\\
         && \mathcal{H}^{alt}(-2.5\leq \lambda \leq 1.0, \Delta=-1.0, B=0.25),\, t>0.
    \end{eqnarray}
    In Fig. \ref{haldane_t_delta} (c), time increment is taken as \(1\) while quenching values of \(\lambda\) are taken with increment \(0.05\). In this situation, we find that at almost \(86.85\%\) points,   GGM goes beyond \(0.49\), thereby demonstrating maximal genuine multipartite entanglement creation in dynamics.

    \item  
    If we only change \(\lambda\) values by fixing both \(\Delta\) and \(B\) at \(t>0\), patterns of entanglement changes drastically. Specifically, for any choice of \(\lambda\) values between \(-2.5\) and \(1.0\), i.e., if the system evolves according to \(\mathcal{H}^{alt}(-2.5 \leq \lambda \leq 1.0, \Delta =1.0, B=0.25)\), there is a range of \(-0.5<\lambda<0\) where \(\mathcal{G}(|\psi_t\rangle)\) vanishes \(\forall t\) while far from that quench, GGM can be created and the maximal amount is also high \(\approx 0.42 \) as shown in Fig. \ref{haldane_t_delta}(d).   
    
    \item The above observations are tempted us to study the behavior of entanglement when the state is initially prepared in the Haldane phase with \(\mathcal{H}^{alt}(\lambda_0 =-0.5, \Delta_0 =1.0,  B_0=0.25)\) and \(t>0\), both \(\lambda\) and \(\Delta\) values are quenched. Specifically, at \(t>0\), when \(-2.5\leq \lambda \leq 1.0\) and \(-1.5\leq \Delta\leq 3.0\), contours of GGM is depicted in Fig. \ref{haldane_t_delta}(a) and (b)  with two different fixed times, \(t=5.0\) and \(t=200\). It again confirms that there is always a \(\lambda, \Delta\)-pair  in which the creation of multipartite entanglement reaches its maximal value. 
     For example, among the total number of \(\approx 6500\) quenching points  generated to create Fig. \ref{haldane_t_delta} (a) and (b), we find \(7.32\%\) of states having GGM more than \(0.4\) at \(t =5.0\) while it is \(12.53\%\) at \(t=200.0\). The percentage decreases with the increase of GGM values although there are some quenching points in (\(\lambda, \Delta\))-plane in which GGM can go beyond \(0.49\). 
    Notice that although the entire study is performed with a fixed initial state, the behavior remains same if the initial state is chosen from other parameter set from the Haldane as well as from the singlet-dimer phase. Specifically, the percentage gets increased if the initial state is the ground state of the singlet phase.

\end{enumerate}

After the above investigations, we can safely conclude that if the Hamiltonian evolution takes place according to the following rule:
\begin{eqnarray}
    && \mathcal{H}_0 \equiv \mathcal{H}^{alt}(-2.5\leq \lambda_0 \leq 1.0, \Delta_0= 1.0, B_0=0.25),\, t=0,\nonumber \\
   && \mathcal{H}_1 \equiv \mathcal{H}^{alt}(-2.5\leq \lambda\leq 1.0, \Delta \neq 1.0, B=0.25),\, t>0, \nonumber\\
\end{eqnarray}
the genuine multipartite entanglement can be created in any time period of the dynamics, thereby establishing AF-F alternating chain as well as AF chain with alternating bonds as potential entanglement resource.





\section{Detecting dynamical quantum phase transition via multipartite entanglement}
\label{sec:dqpt}

To describe equilibrium physics, several physical quantities like partition function, free energy, correlation length, and in recent times, entanglement are identified which can be used to understand the phenomena like quantum phase transition, thermal properties of the system \cite{qptbook1, qptbook2}. On the other hand, while describing dynamical states, no such universal quantities were known. Recently, it was found that if the system is initially prepared in the ground state of a phase and  is then suddenly quenched to another phase for a later time,  there exist certain physical quantities like rate function based on Loschmidt echo, fluctuations in multipartite entanglement are nonanalytic with time in the transient regime -- a phenomena is termed as dynamical quantum phase transition  and the times in which nonanalyticities are observed is known as critical times. In this work, we argue that multipartite entanglement quantifiers, GGM  is as  good as the rate function  to identify DQPT. 

Let us define the function, known as Loschmidt echo (LE), as
\begin{equation}\label{lcho_def}
	\mathcal{L}(t) = |\langle\psi_0|\exp(-i\mathcal{H}^{alt}t|\psi_0\rangle|^2. 
	\end{equation}
Based on it, we can define the rate function, given by \(\mathcal{R}(t) = Lt_{N \rightarrow \infty} \frac{1}{N} \log \mathcal{L}(t)\) which can show nonanlyticities with time, thereby determining the DQPT. Notice that the rate function is analogous  to the partition function in the equilibrium physics \cite{Heylrev} provided we replace $\beta$ in the latter with the complex parameter  $it$.


When the evolution is governed by the transverse Ising model, the rate function clearly shows nonanalyticity at critical times, \(t^* = k(n+1/2),\, n=1,2,3,..\) although  nonuniform critical times are also observed for the XY model with uniform as well as alternating magnetic fields \cite{stavPRB, dqpt_entanglement1,dqpt_entanglement2}. Some of us has also recently argued that just like the rate function,  fluctuations in multipartite entanglement can also detect DQPT \cite{stavPRB} (cf. \cite{stavPRR}).

Before studying the critical times in the entire phase diagram, let us first illustrate the behavior of LE and GGM with time when the initial state is the ground state of the Haldane phase with $\mathcal{H}^{alt}(\lambda_0=-1.00, \Delta_0 = 1.0, B_0=0.25)$. The sudden quench takes place in the singlet-dimer phase, i.e. we choose at \(t>0\), $\mathcal{H}^{alt}(\lambda>0.75, \Delta = 1.0, B=0.25)$. Note here that the initial state is the ground state of the AF-F alternating spin chain while the quenching parameter belongs to the AF chain with alternating bonds.  Unlike the fluctuations in GGM  for the $XY$ model with uniform and alternating magnetic fields,  we report here that the GGM itself of the evolved state can show sharp kinks in the transient time, thereby clearly manifesting nonanalyticity with time. On the other hand,  LE goes to vanish at certain critical times which are responsible for nonanalytic behavior of the rate function.  Note that  we call \(\mathcal{L}(t) =0\) when we find numerically \(\mathcal{L}(t) < 10^{-2}\).

\begin{figure}
		\centering
		\includegraphics[width=9.5cm]{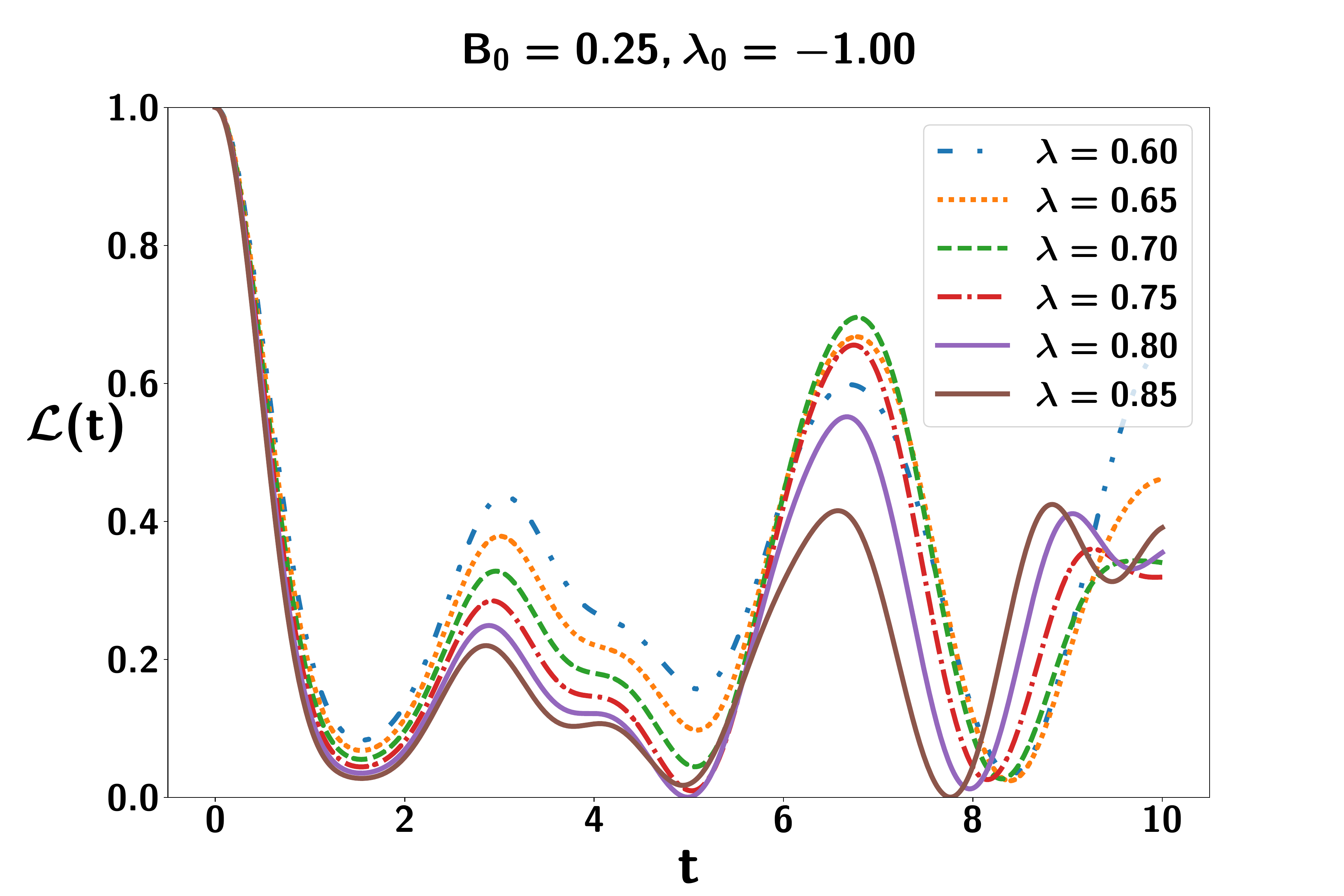}
		\caption{ Loschmidt echo, $\mathcal{L}(t)$ (vertical axis) vs. time, $t$ (horizontal axis) for different $\lambda$ values at \(t>0\). The initial state is the ground state of  \(\mathcal{H}^{alt}(\lambda_0 = -1.0, B_0=0.25, \Delta_0 =1.0)\).  We observe the $t$ values where  $\mathcal{L}(t)=0$ which  corresponds to the non-analytic behavior in the rate function, \(\mathcal{R}(t)\).  Different textures of plots correspond to different $\lambda$ values. Here \(N=12\).    Both the axes are dimensionless.}
		\label{L_echo_quench}
	\end{figure}
\begin{figure}
		\centering
		\includegraphics[width=9.5cm]{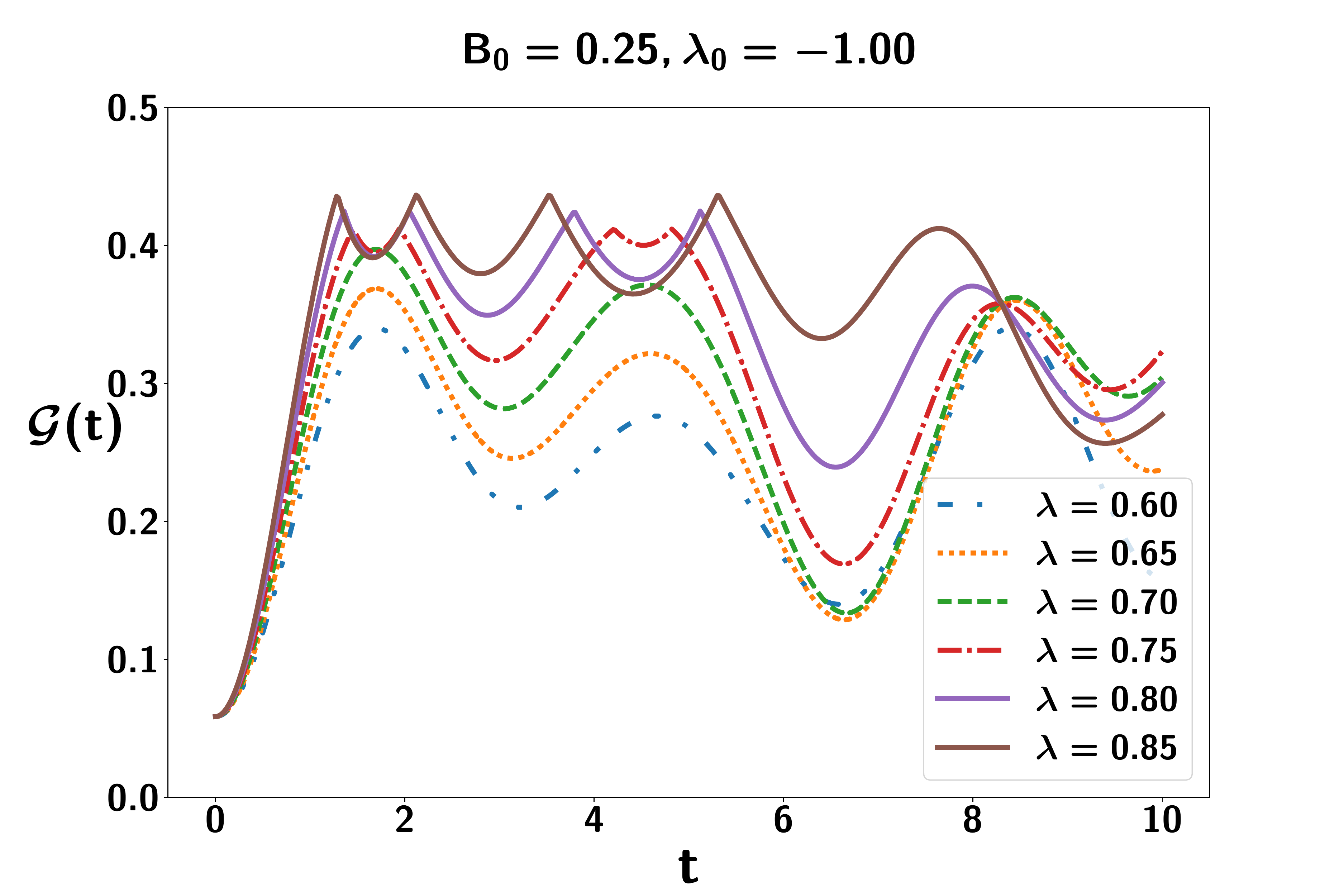}
		\caption{GGM $\mathcal{G}(t)$ (ordinate) against time, $t$ (abscissa) for different $\lambda$ values. All other specifications are same as in Fig. \ref{L_echo_quench}. Both the axes are dimensionless. }
		\label{ggm_quench}
	\end{figure}


 \begin{figure*}
		\centering
		\includegraphics[width=\linewidth]{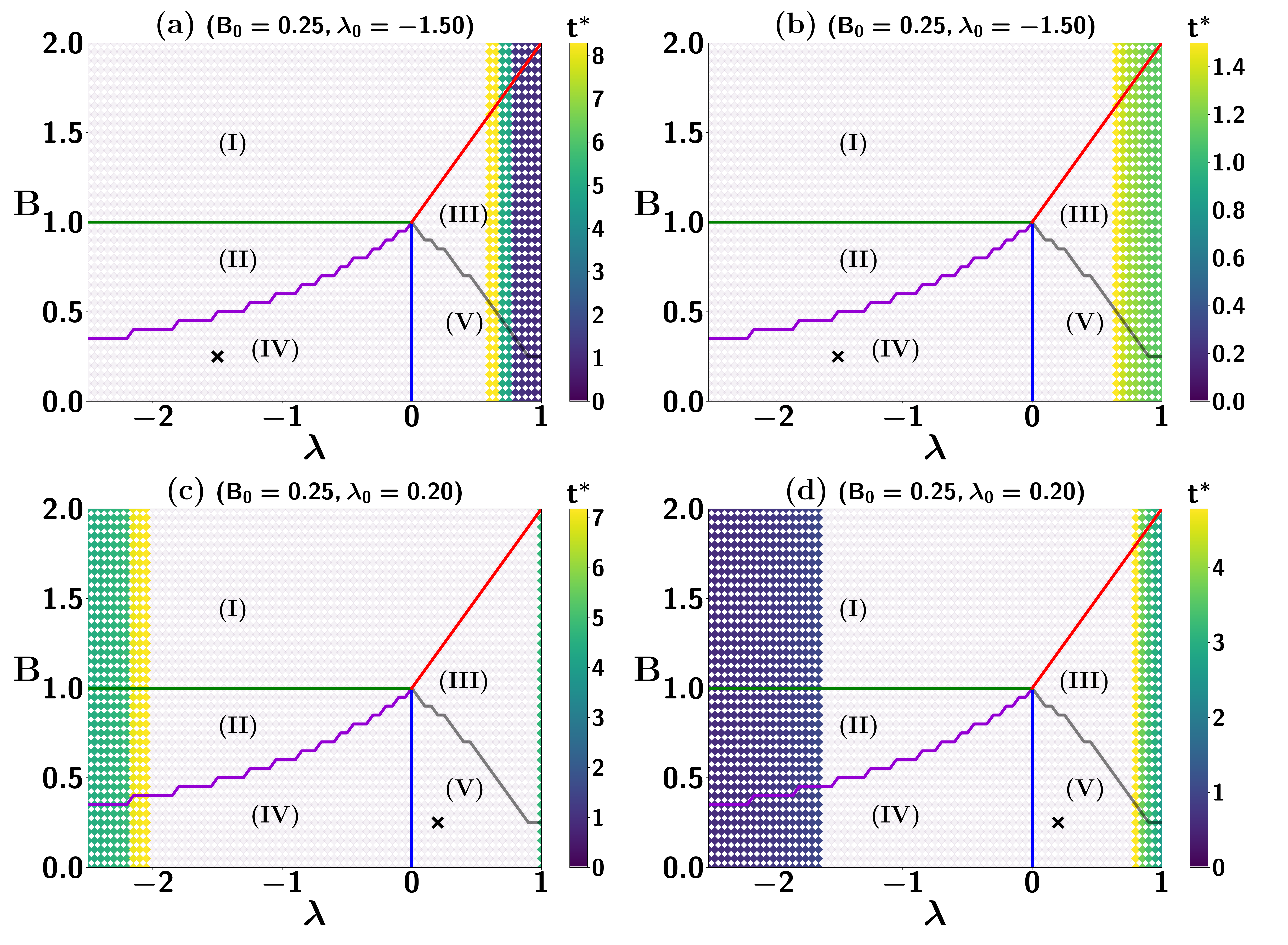}
		\caption{\textbf{Critical times.} Map plot of the critical times obtained from the nonanalyticities of the rate function, \(\mathcal{R}(t)\) ((a) and (c)) and GGM  ((b) and (d)) with respect to the quench in the \(\lambda-B\)- plane. The parameters for the initial states are mentioned in the title of each plot and marked as cross in the figure. The values of \(t^*\) are  mentioned in color palette.  The system-size is considered to be \(N=12\). the  All the axes are dimensionless. }
		\label{dqpt_fail}
	\end{figure*}
 
\subsubsection{Entanglement measures as indicators of DQPT}

 We now argue that multipartite entanglement measures have potential to act as a detector for DQPT like Loschmidt echo. Specifically, we investigate whether first critical time obtained from the  non-analyticity in GGM, \(\mathcal{G}(t)\),  match with those found from LE (see Fig. \ref{dqpt_fail}). To study it, we prepare initial states as the ground states with three sets of parameters, two from the Haldane phase, marked it as (A) and (B) and third one is from the singlet phase, named it as (C). The sudden quench is performed in all other points in the phase diagram by varying \(\lambda\) and \(B\) and by keeping \(\Delta =1.0\) fixed with the initial point.   The time range considered for study is from $t \in \{0,200\}$, the  values of critical times are represented in the color palette and if the critical value is not found in that range, we just mark it as white. We observe a clear distinction between the power of LE and GGM depending on the initial points. 
 
      \textbf{Case (A).} Both LE and GGM can identify the regions when the quench is performed in phases other than Haldane as shown in Figs. \ref{dqpt_fail}(a) and (b).
     The results indicate as shown in the literature that  non-analytic behavior by either rate function or GGM can only guarantee the corresponding phase transition in the equilibrium and not other way round, thereby providing the sufficient condition. 
     
     \textbf{Case (B) and Case (C). } In both the scenarios, LE and GGM can identify the DQPT although they both can show anomalous behavior, eg.  when \(\lambda \approx 1\)  as shown in Figs. \ref{dqpt_fail}(c) and (d). It is not clear whether this is due to the calculations performed with finite system-size  or it wrongly signals  the equilibrium transition. Furthermore, in dynamics, the evolution does not take place by changing \(B_0\) to \(B\) although there are critical lines which appear via tuning of \(B\) and hence this model is different than the one considered before in the context of DQPT.   

	\begin{figure*}[ht]
		\centering
		\includegraphics[width=\linewidth]{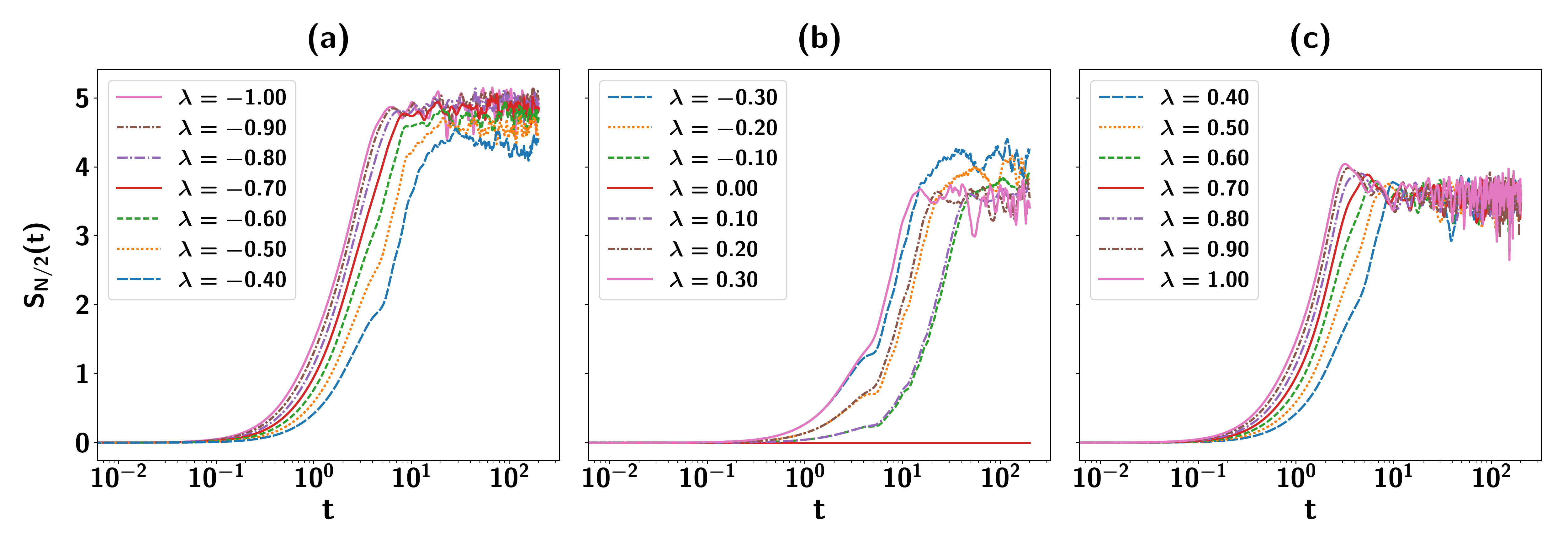}
		\caption{\textbf{Half entanglement-entropy: a detector of quantum criticalities in equilibrium.} (a) Entanglement entropy, \(S_{N/2}(t)\) (vertical axis) vs. \(t\) (horizontal axis) when the initial state is the N{\'e}el state and the evolution is governed by the  \(\mathcal{H}^{alt}(\lambda, \Delta =1.0, B= 0.25)\) where the values of \(\lambda\) are mentioned in the legend. They correspond to  the  Haldane phase in equilibrium.  (b) and (c) Same as in (a), except the values of \(\lambda\) in (b) signifies the neighborhood of the critical lines  while in (c), they are chosen from the singlet-dimer phase. Here \(N=12\).
		All the axes are dimensionless. }
		\label{half_entropy}
	\end{figure*}


\subsection{Identification of quantum phases from dynamics of entanglement entropy  }
\label{subsec:entent}

 Until now, we consider the ground state of a model as the initial state for evolution. Let us now start with a N{\'e}el state as the initial state, i.e., $|\phi_0 \rangle = |\uparrow \downarrow \uparrow \downarrow \cdots \rangle$.  We  evolve the system according to a Hamiltonian \(\mathcal{H}^{alt}(\lambda, \Delta, B)\) with  different set of parameters so that they correspond to different phases at equilibrium. The evolved state in this case reads as
 \begin{eqnarray}
     |\phi_t \rangle = \exp(-i\mathcal{H}^{alt}(\lambda, \Delta, B)t) |\phi_0 \rangle.
 \end{eqnarray}
 The proposal is as follows:\\
 \textbf{The dynamics of multipartite entanglement quantified via the half-block entropy of the evolved state can predict the equilibrium phase transition in the transient regime.}\\
The half-block entropy is the von Neumann entropy of the evolved state after partitioning it equally, i.e., we compute \(S_{N/2}(t) \equiv S( |\psi_t \rangle_{N/2: N/2})\) with \(S(\sigma) = - \mbox{tr} (\sigma \log_2 \sigma)\). 

 To establish the above proposition, we divide the parameter regimes in three distinct parts for constructing the evolution operator -- (A)  \(\lambda\) values are chosen in the Haldane phase but away from the boundary of Haldane and singlet-dimer phase, i.e.,  $\lambda << 0$, (B) around the boundary, i.e., $\lambda \approx 0$ and (C) \(\lambda\) is deep into the singlet-dimer phase, i.e., $0.4 \leq \lambda \leq 1.0 $. \\
 
     \textbf{Regions (A) and (C).} The evolution of $S_{N/2}(t)$ for different values of  $\lambda$  is almost identical with the variation of \(t\), thereby making a band-like structure with time  in the transient regime as shown in Fig. \ref{half_entropy}. Moreover, we notice that in the steady state, $S_{N/2}(t)$ converges to a higher value when the system parameters in the evolution operator are chosen from the Haldane phase compared to the  situation when they belong to the singlet-dimer phase in equilibrium (comparing Figs. \ref{half_entropy}(a) and (c)). 
     
      \textbf{Region (B).} When the system evolves according to the Hamiltonian close to the critical lines, $S_{N/2}(t)$ becomes distinguishable with \(t\) for different choices of \(\lambda \approx 0\) and the band size gets increased in this case (see fig. \ref{half_entropy}).  
     

     The entire analysis reveals that starting from a suitable product state, the band-width quantifying the difference between the values of the block-entropies with time for different system parameters can predict whether during the evolution,  parameters chosen correspond to far from  quantum phase transition at equilibrium   or to the quantum criticalities. Specifically, the band-width increases when the system parameters are chosen close to the quantum critical lines.  
     
 

	\section{Conclusion}
	\label{sec:conclu}
	
 Discovery of exotic phases, and counter-intuitive phenomena make the study of quantum spin models attractive.   In recent times, these models can also be realized and controlled in laboratories by using ultracold atoms in optical lattices, trapped ions, superconducting qubits, thereby raising possibilities to probe the system. We consider a one-dimensional $XXZ$ model with alternating bond strength which can be ferromagnetic (F) or antiferromagnetic (AF) in presence of magnetic field. It was  shown that the spin-1/2 AF-F alternating chain with magnetic field  can possess certain phases which were conjectured to be only present in  models with integer spins. Although the model cannot be solved analytically, the phase diagram is well understood via the numerical and approximate techniques.  
	
	In this work, we  addressed whether the properties of the dynamical state of the system can mimic faithfully the ground state phase diagram or not. Towards characterizing non equilibrium physics, we first showed that the trends of  both bipartite as well as multipartite entanglement can infer   quantum criticalities at zero temperature in the system, thereby reproducing the rich phase diagram of the $XXZ$  chain with alternating bond strengths. The entire analysis is carried out by using Lanczos algorithm with a spin chain consisting of sixteen or twelve spins depending on the static or dynamical scenarios.


Starting from the ground state, we found that high multipartite entangled states can be created in this model by suitably choosing the initial state and  parameters in sudden quench, thereby establishing  it as a good  entanglement resource. We then observed that the phase transition at zero temperature from the Haldane to the singlet-dimer phase can be identified by studying the dynamics of the multipartite entanglement which shows nonanalyticity with time provided the initial and the evolved state-parameters belong to two different  phases. On the other hand, when a product state evolves to an entangled state governed by the AF-F alternating spin chain or AF spin chain with alternating bond strength, the dynamics of block entropy with time can also indicate whether the Hamiltonian corresponding to the evolution operator is close to the phase transition or not. 

The importance of the $XXZ$ alternating spin chain  is that even when the spins are half-integer,  due to the competition between the interaction strengths in the alternating bonds, it can mimic certain properties  which are typically present only in higher dimensional systems. The exotic phases turn out to be responsible for high entanglement generation in dynamics, thereby indicating that it has potential to  design  quantum  devices.


	\acknowledgements
	The authors acknowledge the support from  the Interdisciplinary Cyber Physical Systems (ICPS) program of the Department of Science and Technology (DST), India, Grant No.: DST/ICPS/QuST/Theme- 1/2019/23.  We  acknowledge the use of the cluster computing facility at the Harish-Chandra Research Institute.

	\bibliography{axyz.bib}
\end{document}